\documentclass{PoS}

\def\ltsima{$\; \buildrel < \over \sim \;$}
\def\simlt{\lower.5ex\hbox{\ltsima}} % < over ~
\def\gtsima{$\; \buildrel > \over \sim \;$}
\def\simgt{\lower.5ex\hbox{\gtsima}} % > over ~

\def\deg{\hbox{$^\circ$}}

\def\gray{$\gamma$-ray}
\def\grays{$\gamma$-rays}

\def\r95{$r_{\rm 95}$}
\def\p0{$\pi^{\rm 0}$}
\def\t0{$t_{\rm 0}$}

\def\Compton{\textit{Compton}}
\def\Fermi{\textit{Fermi}}

\def\Suzaku{\textit{Suzaku}}

\def\ncyg{{\rm V407 Cyg}}
\def\nsco{{\rm V1324 Sco}}
\def\nmon{{\rm V959 Mon}}
\def\ndel{{\rm V339 Del}}
\def\ncen{{\rm V1369 Cen}}
\def\nsgr{{\rm Nova Sgr 2015 No.~2}}
\def\nvel{{\rm V382 Vel}}

\def\aap{{\it Astron.~Astrophys.}}

\def\aj{{\it Astron.~J.}}

\def\aph{{\it Astropart.~Phys.}}
\def\apj{{\it Astrophys.~J.}}
\def\apjl{{\it Astrophys.~J.~Lett.}}
\def\apjs{{\it Astrophys.~J.~Supp.}}

\def\atel{{\it The Astronomer's Telegram}}

\def\iaucirc{{\it IAU Circ.}}
\def\mnras{{\it Mon.~Not.~R.~Astron.~Soc.}}
\def\nat{{\it Nature}}

\def\prd{{\it Phys.~Rev.~D}}
\def\procspie{{\it Proc.~SPIE}}

\def\sci{{\it Science}}

\def\etal{{\it et al.}}

\usepackage{lineno}
\usepackage{rotating}
\usepackage{color}
\title{\Fermi\ Reveals New Light on Novae in Gamma rays}

\ShortTitle{Gamma-Ray Novae}

\author{\speaker{C. C. Cheung}\\
        Naval Research Laboratory, USA\\
        E-mail: \email{Teddy.Cheung@nrl.navy.mil}}

\author{P. Jean\\
        IRAP \& Universit\'e de Toulouse, France\\
        E-mail: \email{Pierre.Jean@irap.omp.eu}}

\author{S. N. Shore\\
        University of Pisa, INFN, Italy\\
        E-mail: \email{shore@df.unipi.it}}

\author{J. E. Grove\\
        Naval Research Laboratory, USA\\
        E-mail: \email{Eric.Grove@nrl.navy.mil}}

\author{M. Leising\\
	Clemson University, USA\\
        E-mail: \email{lmark@clemson.edu}}

\author{on behalf of the \Fermi-LAT Collaboration}

\abstract{Novae are now firmly established as a high-energy ($>$100 MeV) \gray\ source class by the \Fermi\ Large Area Telescope (LAT). In symbiotic binary systems such as V407 Cyg 2010, there is a firm theoretical framework for the production of shock-accelerated particles in the nova ejecta from interactions with the dense wind of the red giant companion. Yet, the high-energy \gray\ emission detected in classical novae involving less evolved stellar companions cannot be explained in the same way and could instead be produced in internal shocks in the ejecta. We summarize the \Fermi-LAT \gray\ observations of novae, highlighting the main properties that will guide further studies. Additionally, we report on the soft \gray\ ($\sim 0.1$ MeV) continuum detection of the oxygen-neon type classical nova V382 Vel 1999 with the OSSE detector aboard the \Compton\ Gamma Ray Observatory in light of its \Fermi-era analog, V959 Mon 2012.}

\FullConference{The 34th International Cosmic Ray Conference,\\
		30 July - 6 August, 2015\\
		The Hague, The Netherlands}

\begin{document}

%%%%%%%%
%% Begin
%%%%%%%%

\section{\Fermi\ LAT Gamma-ray Novae Discoveries}

Novae were not among the wide array of high-energy \gray\ ($>$100 MeV) source types considered prior to the launch of the \Fermi\ Large Area Telescope (LAT) \cite{atw09}. That changed when the symbiotic-like recurrent nova \ncyg\ 2010 explosion was discovered  optically and identified with an independently detected \Fermi-LAT transient source \cite{nis10,abd10,che10}. In retrospect, the idea that such symbiotic systems can be high-energy \gray\ emitters could have been anticipated as a product of shock-accelerated particles in the nova ejecta through interactions with the dense wind of the red giant companion as considered for RS Oph \cite{tat07}. However, the later LAT detections of three classical novae (CNe) were truly unexpected \cite{ack14} (see Figure~1) because they were produced in compact binary systems composed of a white dwarf and main sequence companion. The detected CNe also had different optical/ultraviolet properties, with \nmon\ identified as an explosion from an oxygen-neon (ONe) nova, and \nsco\ and \ndel\ appear to be carbon-oxygen (CO) novae.  

Despite the compositional differences, the \gray\ light curves for the ensemble four systems appear similar, being brighter initially with most of the emission observed within $\sim2$ weeks, and total detected durations of $17-27$ days (Figure~2) \cite{ack14}. Because \nmon\ was discovered first by \Fermi-LAT as a \gray\ transient in the Galactic plane \cite{che12mon1}, then identified as a nova $\sim2$ months later following its optical discovery \cite{fuj12,che12mon2,che12}, a comparison of the optical and \gray\ light curves is not possible. For the other three systems (Figure~3; see \cite{abd10} for \ncyg), the \gray\ peaks lagged the optical peaks (by up to $\sim$ 6 days in the case of \ndel), with the first significant LAT detection occurring before (\nsco), after (\ndel\footnote{But note a $2.4\sigma$ detection on the day of the optical peak in a LAT 0.5 day binned light curve \cite{ack14}.}) and on the same day (\ncyg) of the observed optical peak. \nsco\ was unusual in that its optical activity began about 1-month prior to its first LAT detection on 15 June, having displayed an increase of 6 mag between June $1.77-3.33$ after an initial discovery of an enhanced optical flux in mid-May \cite{wag12,mun15}.

The observed $>$100 MeV spectra of the four novae are also similar, with broad spectral peaks that cutoff at energies $\sim 1-4$ GeV and emission observed up to $\sim 6 - 10$ GeV (the highest energy extension was observed in \nsco). Spectral fits with hadronic and leptonic models were indistinguishable (e.g., Figure~4 [left]). The available distance estimates to the novae of $\simlt 4-5$ kpc indicate they are nearby, and the resultant range of $>$100 MeV luminosities and total emitted energies spanned a small range of $\sim (3-4) \times 10^{35}$ ergs s$^{-1}$ and $\sim (6-7) \times 10^{41}$ ergs, respectively, with $\sim2\times$ larger values for \nsco. The \ndel\ distance of 4.2 kpc in \cite{ack14} was confirmed \cite{sch14} and the 2.7 kpc distance for \ncyg\ is unchanged. For \nmon\ (distance of 3.6 kpc; \cite{sho13}), adopting smaller inferred distances of $\sim 1.5$ to 2.2 kpc \cite{mun13,lin15} would translate to a \gray\ luminosity $\sim5\times$ to $3\times$ smaller.  As cautioned for \nsco\ (see SOM of \cite{ack14}), the adopted 4.5 kpc distance \cite{hil12} could be underestimated due to dust and has a likely range of $4-8$ kpc, consistent with independent estimates \cite{mun15,fin15} so its luminosity value could be even higher. To what extent the \gray\ parameters (luminosities, spectra, and durations) of the LAT-detected novae are similar is not yet understood, but this can assuredly be addressed with future observations (\S~4).

\begin{figure}[t]
%\vspace{-0.1in}
%\hspace{0.65in}
\centering
\includegraphics[height=0.55\textheight]{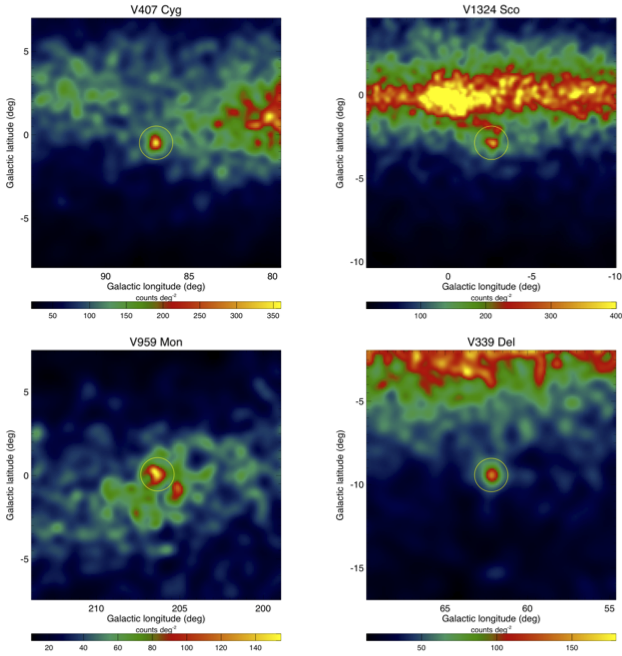}
%\vspace{-0.1in}
     \caption{\Fermi-LAT adaptively smoothed $>$100 MeV counts maps of the first four \gray\ detected novae, \ncyg\ 2010, \nsco\ 2012, \nmon\ 2012, and \ndel\ 2013. Each nova is placed at the map centers (marked with yellow 1\deg\ radius circles) and observed near bright diffuse Galactic \gray\ emission (\nmon\ is seen directly through the plane). This Figure is taken directly from \cite{ack14}.}
     \label{fig1}
\end{figure}

\begin{figure}[t]
%\hspace{2.5cm}
\centering
\includegraphics[height=0.45\textheight]{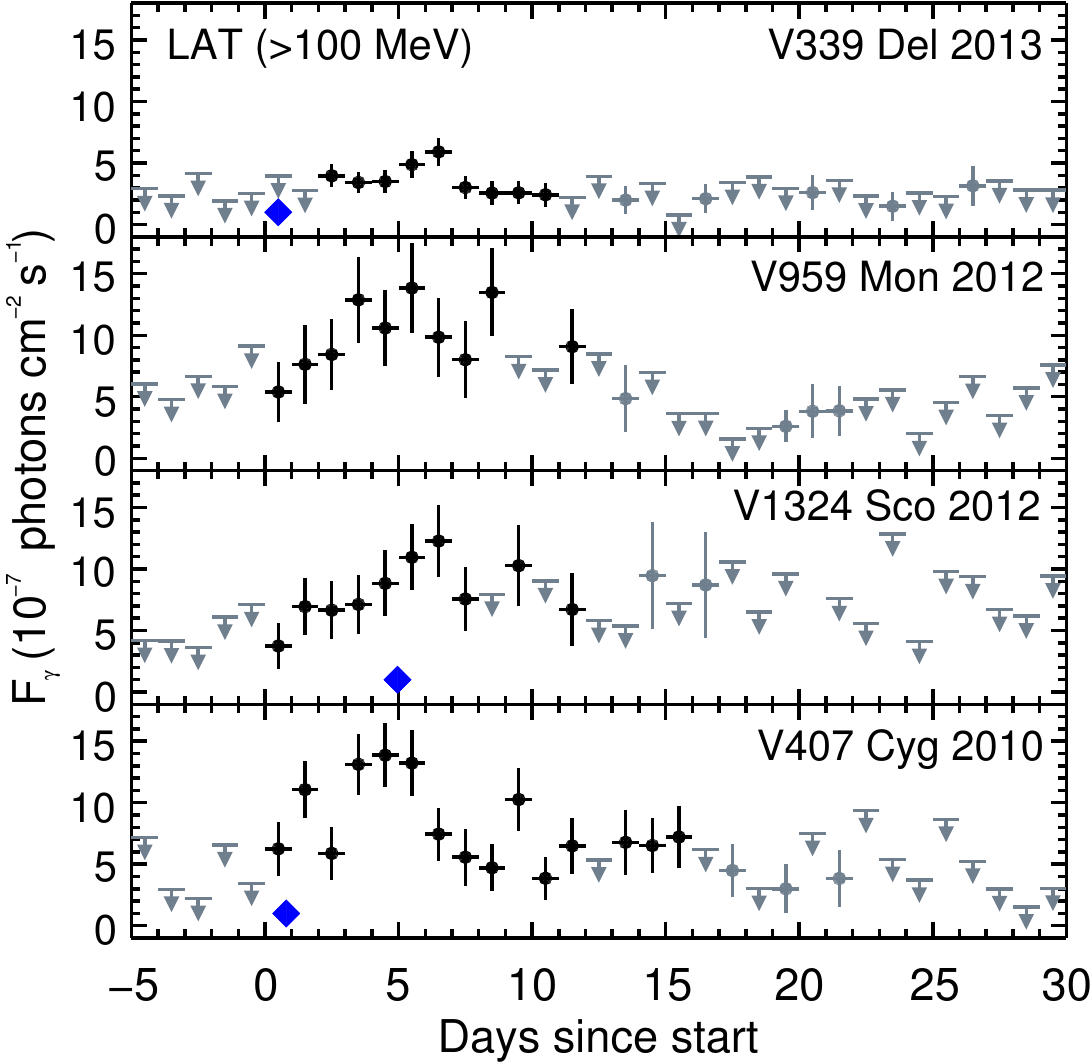}
%\vspace{-0.1in}
     \caption{\Fermi-LAT $>$100 MeV 1-day binned light curves of the the first four \gray\ detected novae (cf., Figure~1). Vertical bars indicate 1$\sigma$ uncertainties for $>3\sigma$ points (black) and $2$-$3\sigma$ (gray) significances, otherwise, 2$\sigma$ upper limits are indicated with gray arrows. Start times (from top to bottom panels) of 2013 August 16, 2012 June 19, 2012 June 15, and 2010 March 10 were defined as the day of the first LAT detection (there was a $2.4\sigma$ detection of \ndel\ in 0.5-day binned data from August 16.5-17.0; see \cite{ack14}). The epochs of the observed optical peaks are marked with a blue diamond in each panel except for \nmon, where the optical peak was unobserved due to proximity to the Sun. This Figure is taken directly from \cite{ack14}.}
     \label{fig2}
%\vspace{-0.2in}
\end{figure}

\begin{figure}[t]
%\vspace{0.3in}
\centering
     \includegraphics[height=0.3\textheight]{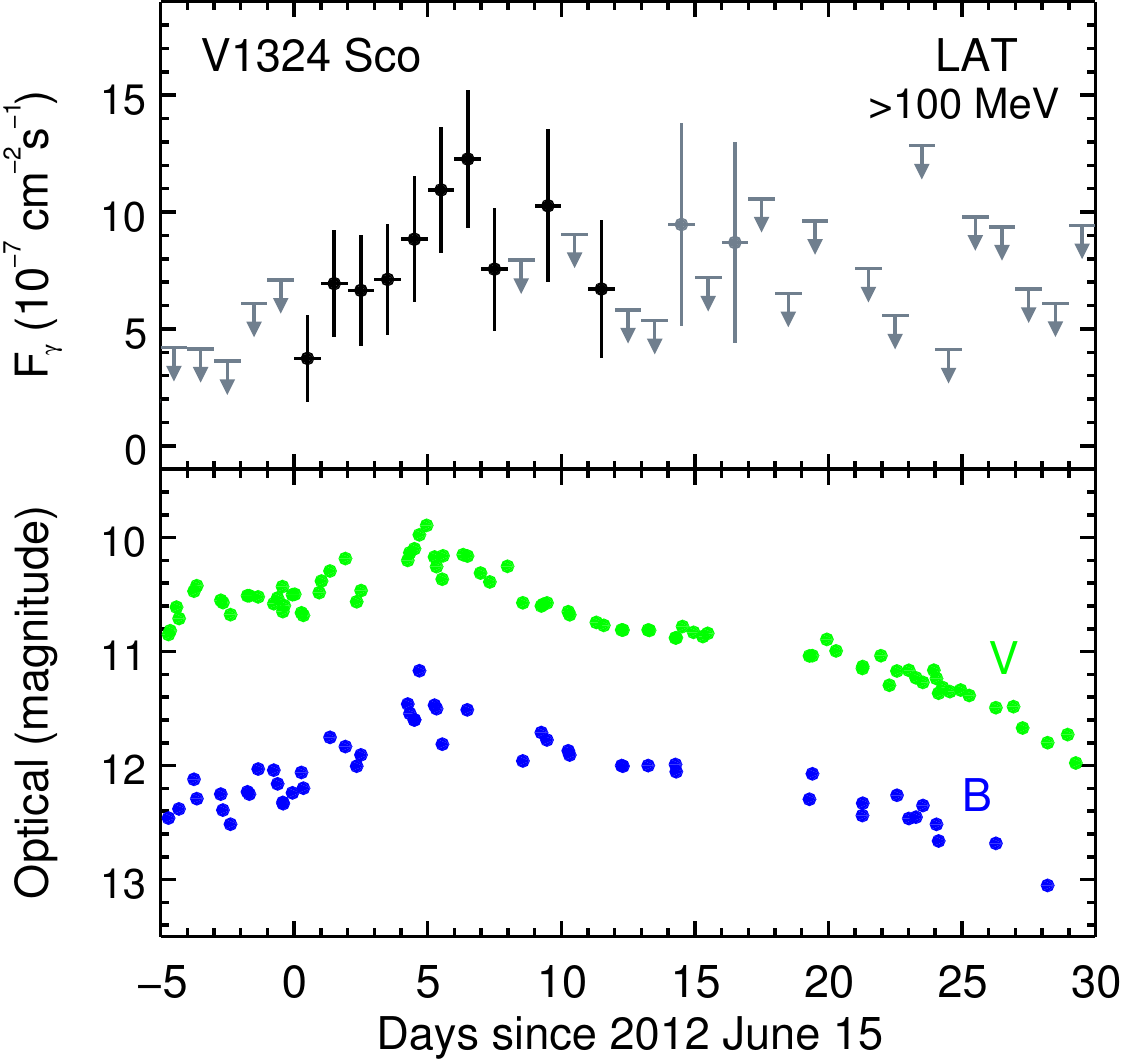}\includegraphics[height=0.3\textheight]{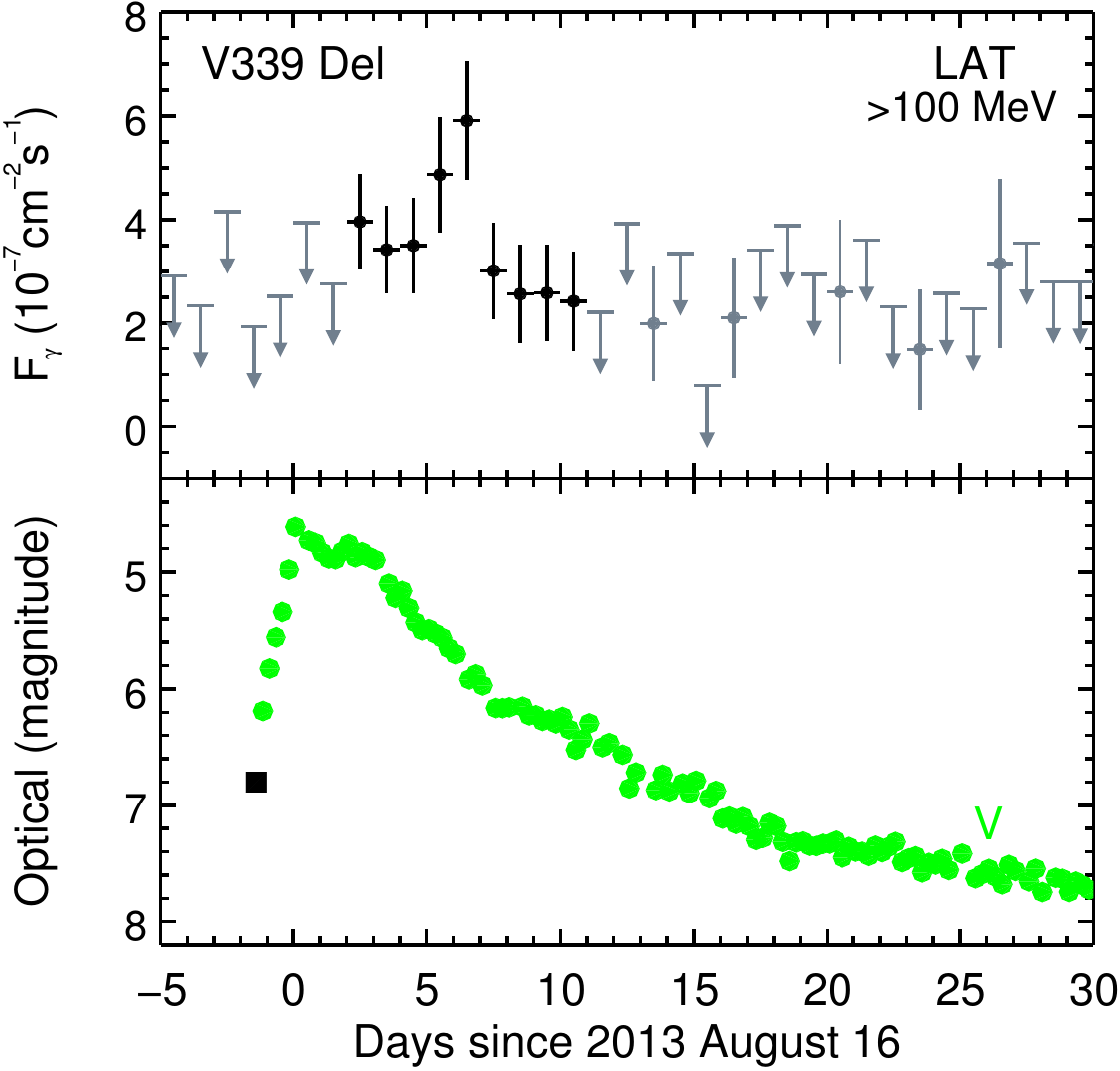}
%\vspace{-0.8cm}
     \caption{\Fermi-LAT $>$100 MeV 1-day binned light curves (top panels) of \nsco\ and \ndel\ as shown in Figure~2, with optical light curves (bottom panels) over the same time interval.  For \nsco,  the optical data are those compiled by \cite{mun15} predominantly from the AAVSO. The \ndel\ figure is taken directly from \cite{ack14} where the optical data are from the AAVSO in 0.25 day bins and includes the additional unfiltered discovery magnitude (black square) \cite{ita13}; note the small second peak $\sim$2 days after the first.}
     \label{fig3}
\end{figure}

%%%%%%%%%
%% OSSE Vel

\section{\Compton\ OSSE Continuum Detection of V382 Vel 1999}

The \Fermi-LAT $>$100 MeV classical novae detections motivated us to revisit previous observations of other novae from the \Compton\ Gamma Ray Observatory (CGRO). In the case of the bright classical nova V382 Velorum 1999 (\nvel), that peaked at $\sim$2.5 mag about one-day after it was discovered May 22.4 UT \cite{wil99}, Oriented Scintillation Spectrometer Experiment  (OSSE) \cite{joh93} observations were obtained over 26 days beginning May 27. The primary objective was to search for nuclear line emission, which was not detected (\cite{lei99}; see \cite{her08} for the flux limits for the lines at 0.478 MeV from $^{\rm 7}$Be and 1.275 MeV from $^{\rm 22}$Na). Instead, \cite{lei99} found a $7.0\sigma$ continuum signal over the $0.05-0.3$ MeV energy range in the first 12 days of the OSSE observations (CGRO Viewing Period, VP 817.5), that could not at the time be unambiguously associated with the nova because of OSSE's large field of view ($3.8\deg \times 11.4\deg$). There was no significant detection in the subsequent 14-day observation interval (VP 819.5) that began June 8.

With the benefit of hindsight, these \Compton-OSSE continuum observations of \nvel\ are not surprising considering it is an analogue of the \Fermi-LAT detected ONe-type nova \nmon\ \cite{sho13,che12mon2}.  Specifically, the OSSE soft \gray\ detection within the first $\sim 2$ weeks after its optical peak and its null-detection in the weeks that followed, is similar to the $>$100 MeV light curves of the \Fermi-LAT detected novae (Figure~2), strongly suggesting the positive detection is not a result of source confusion. 

The OSSE $0.05-0.3$ MeV spectrum from VP 817.5 (encompassing the interval $4-16$ days after optical peak) that we can now associate with \nvel\ (no significant detection  was obtained from $0.3-10$ MeV) is compared to the average \Fermi-LAT $>$100 MeV spectrum of \nmon\ 2012 up to 22-days after its first significant 1-day LAT detection (Figure~4 [right]). Because these novae are similarly distant ($1.7-2.5$ kpc for \nvel, \cite{del02,sho03}; see \S~1 for \nmon), this can be considered an early time composite spectrum of an ONe-type nova. Interestingly, a single \textit{BeppoSAX} observation of \nvel\ obtained 15 days after optical maximum resulted in a $0.1-10$ keV detection and $15-60$ keV upper limit ($2\sigma$) $<6.7 \times 10^{-12}$ erg cm$^{-2}$ s$^{-1}$ attributable to a $kT \approx 6$ keV plasma \cite{ori01}. It is difficult to gauge how the \textit{BeppoSAX} data connect to the OSSE spectrum because it probed a single epoch coinciding with the end of the \Compton\ VP (perhaps suggesting fading of the $\sim$0.1 MeV emission), but such early X-ray observations can similarly probe internal shocked emission in the nova ejecta (see \cite{muk01}).

\begin{figure}[t]
\centering
\includegraphics[height=0.3\textheight]{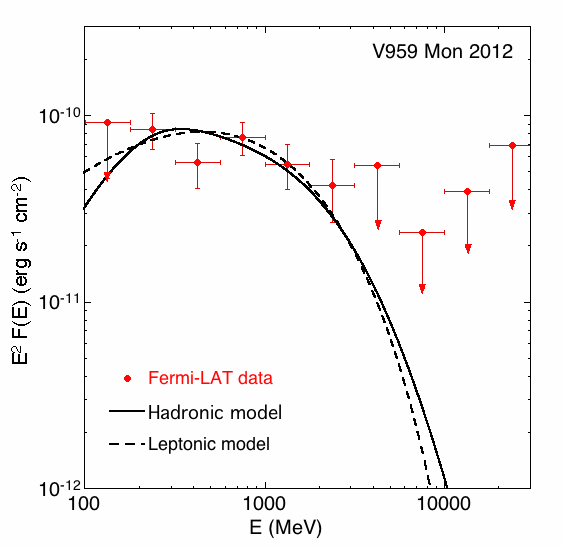}\includegraphics[height=0.287\textheight]{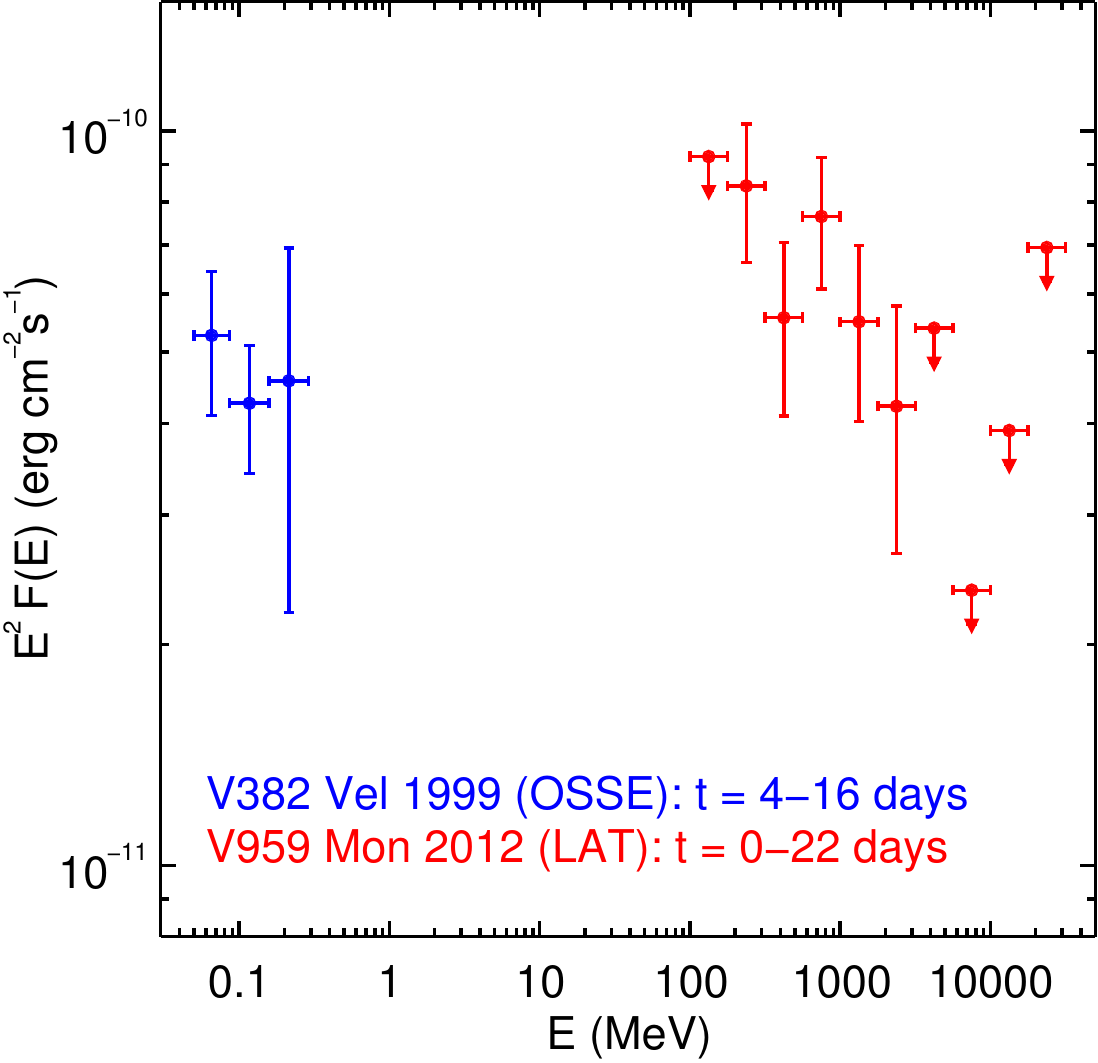}
     \caption{[left] \Fermi-LAT $>$100 MeV average \gray\ spectrum of the classical nova \nmon\ 2012 over 22 days of data. Figure taken directly from \cite{ack14}. Vertical bars indicate 1$\sigma$ uncertainties for data points with significances $>2\sigma$, and arrows indicate 2$\sigma$ limits and the best-fit hadronic (from \p0-decay) and leptonic (from inverse Compton and lower-level bremsstrahlung emission) model curves are overlaid. See \cite{ack14} for details and for the plots for the other two LAT detected classical novae (\nsco, \ndel) and for the symbiotic nova \ncyg\ (see \cite{abd10} as well). [right] A `composite' early-time \gray\ spectrum of an ONe-type classical nova consisting of the \Compton\ OSSE $0.05-0.3$ MeV spectrum from \nvel\ 1999 and the \Fermi-LAT $>$100 MeV spectrum of \nmon\ 2012 from the left panel.}
     \label{fig4}
\end{figure}

%%%%%%%%%%%
%% New Novae

\section{New \Fermi\ LAT Gamma-ray Novae Detections}

Since the first four \gray\ novae discoveries (\S~1), significant LAT detections of two more followed, \ncen\ 2013 \cite{che13a,che13b} and \nsgr\ \cite{che15a,che15b}, as a result of \Fermi\ target-of-opportunity (ToO) observations prompted by their optical discoveries \cite{sea13,sea15}. Both reach naked-eye brightness optically, but in contrast to the previous \gray\ detected novae, several subsequent optical peaks were observed up to $\sim$$1-2$ months after their first optical peaks. The first significant 1-day \gray\ detections began $\sim2$ days after the first optical peaks, similar to that reported in \ndel. The distances to these two novae are not yet known, but they are fainter in \grays\ on average than the previous LAT-detected novae, and are characterized by more sporadic low-significance 1-day detections extended over moderately longer durations (up to $\sim$$1-2$ months), broadly mimicking the multi-peaked optical activity (Figure~5).

\begin{figure}[t]
%\vspace{-0.2cm}
%     \hspace{0.1cm}
\centering
\includegraphics[height=0.3\textheight]{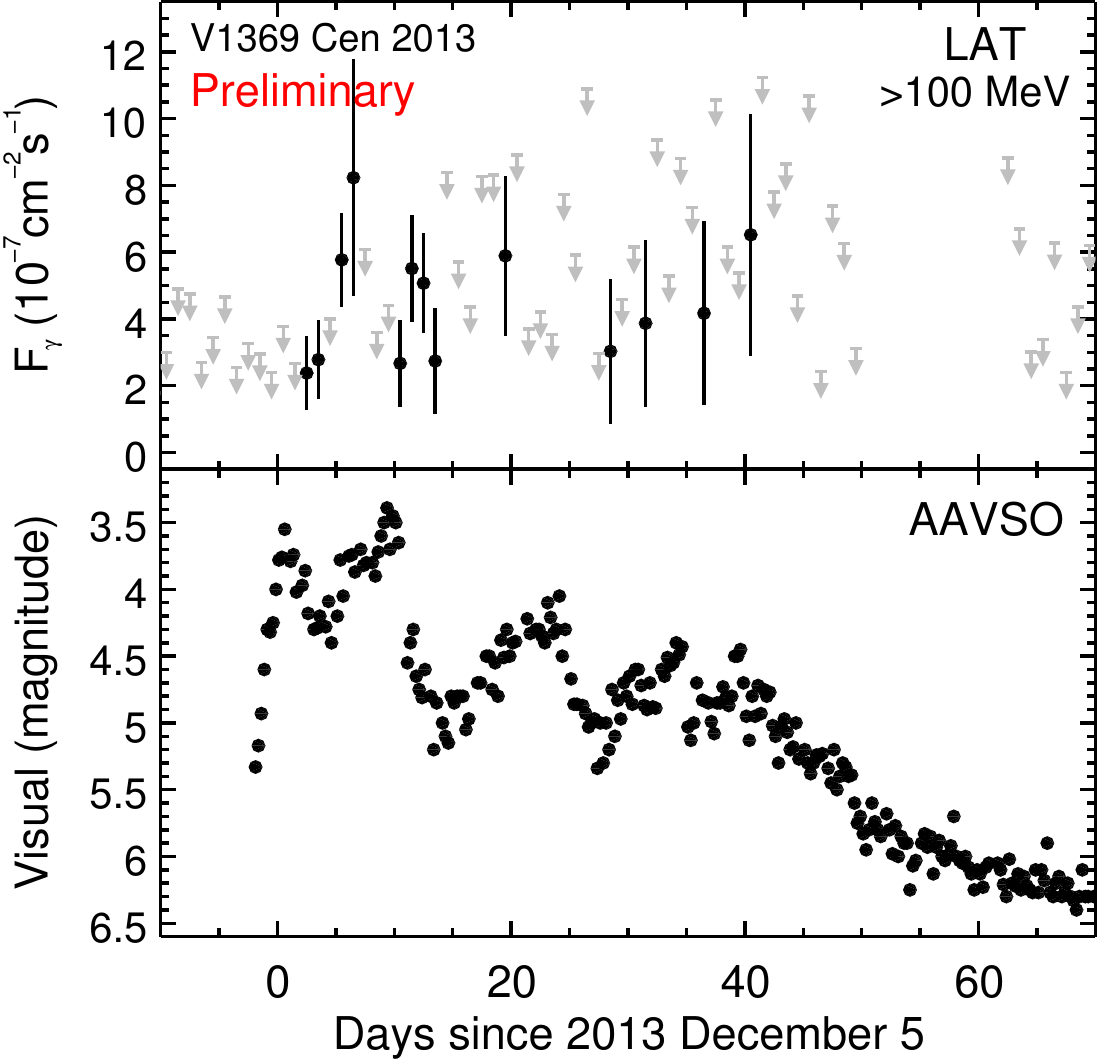}\includegraphics[height=0.3\textheight]{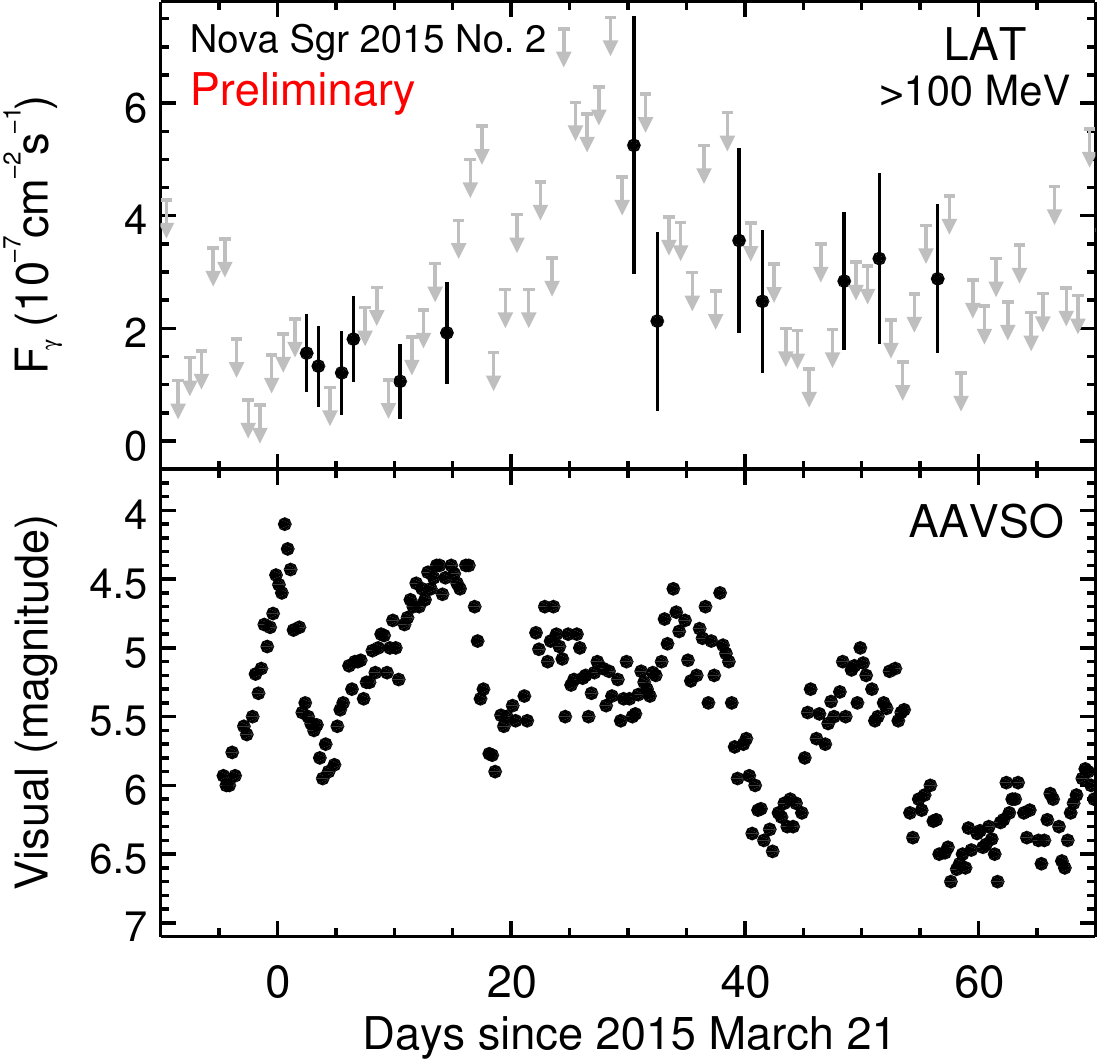}
     \caption{Preliminary \Fermi-LAT $>$100 MeV \gray\ light curves (top panels) of \ncen\ (left) and \nsgr\ (right). Vertical bars indicate $1\sigma$ uncertainties for data points with TS $> 4$ (black), while gray arrows indicate $2\sigma$ upper limits when TS $< 4$. In each case, the \gray\ light curve is compared to the AAVSO Visual light curve in 0.25 day bins for the same time interval (bottom panels).}
     \label{fig5}
\end{figure}

\section{Future Perspectives}

The \gray\ detections of \ncyg\ 2010 and the two CNe in 2012 resulted naturally from \Fermi-LAT's all-sky \gray\ monitoring. They paved the way for the secure detections of fainter \gray\ fluxes from three naked-eye novae enabled by \Fermi\ ToO observations following their optical discoveries by the amateur community. The detection rate of \gray\ novae is $\sim$1 yr$^{-1}$ over the first $\sim$7 yrs of the \Fermi\ mission, consistent with the rate of Galactic novae at $\simlt 4-5$ kpc distances, suggesting all novae are potential \gray\ sources \cite{ack14}. The more recent detections may be revealing a wider diversity in \gray\ properties, being fainter and seemingly characterized by longer total durations.  Further serendipitous and targeted \gray\ detections through the continued \Fermi\ mission, as well as studies of non-LAT detected novae (e.g., \cite{che12}) could reveal the true range of \gray\ properties. Coordinated \gray\ and multi-wavelength observations can test scenarios for how and where the putative shocks are generated (e.g., internal shocks or strong turbulence driven in the ejecta \cite{ack14}, or in wind-wind interactions \cite{cho14}) and point to the underlying \gray\ emission mechanism. 

\Fermi-LAT observations of the recurrent nova V745 Sco 2014 \cite{stu14} indicated low significance, $2-3\sigma$ \gray\ emission on February 6 and 7, coincident with its optical outburst detected on February 6.694 \cite{che14}, suggesting a connection between these events.  Though not definitive, the short \gray\ duration would imply a`fast' analogue of \ncyg\ \cite{ban14}, with the ($\sim3\times$) larger distance of V745 Sco helping to explain the $5\times$ fainter \gray\ peak  flux observed. Such systems demonstrate the important role of the companions in the \gray\ production that has yet been fully explored in CNe. Of the known recurrent novae, one can anticipate future outbursts \cite{sch10} with the immediate ones to watch being perhaps V3890 Sgr (expected in 2015 with some great uncertainty) and U Sco ($2020 \pm 2$ years, again during the \Fermi-era after its 2010 eruption). 

In light of the \Fermi-LAT discovery that classical novae can be bright continuum emitters at $>$100 MeV, the possibility that these spectra extend down to lower, $\sim 0.1$ MeV energies is supported by the revisited \Compton/OSSE observations of \nvel\ 1999. This should motivate early observations with INTEGRAL to search for both soft \gray\ continuum, in addition to nuclear lines, in future \Fermi-novae. Current facilities that can probe the hard X-ray band like \Suzaku, NuSTAR, and the upcoming ASTRO-H mission \cite{tak12} will be invaluable, as well as future MeV missions (e.g., AstroMeV; http://astromev.in2p3.fr). Extrapolating their LAT spectra to higher energies suggest novae could be detected by CTA (see e.g., \cite{bed13} for \ncyg), and can give crucial hints as to the emission mechanism. Simultaneity of such observations will be key, so perhaps then the most important recurrent novae to anticipate for the \Fermi\ mission in the CTA-era are RS Oph (2021? Its last explosion in 2006 did not have \gray\ coverage) and T CrB (2026?), a very nearby $\sim0.8$ kpc system that could produce a remarkably bright MeV-GeV-TeV \gray\ source, and a transient neutrino signal expected in the hadronic scenario \cite{raz10} potentially detectable at even higher energies.

%%%%%%%%
%% All else

\acknowledgments

The \Fermi-LAT Collaboration acknowledges support for LAT development, operation and data analysis from NASA and DOE (United States), CEA/Irfu and IN2P3/CNRS (France), ASI and INFN (Italy), MEXT, KEK, and JAXA (Japan), and the K.A.~Wallenberg Foundation, the Swedish Research Council and the National Space Board (Sweden). Science analysis support in the operations phase from INAF (Italy) and CNES (France) is also gratefully acknowledged. 

C.C.C.~was supported at NRL by a Karles' Fellowship and by NASA through Guest Investigator programs 12-FERMI12-0026 and 13-FERMI13-0008.

We acknowledge with thanks the variable star observations from the AAVSO International Database contributed by observers worldwide and used in this research, and the dedicated observers of the Astronomical Ring for Access to Spectroscopy group (http://www.astrosurf.com/aras/).


\begin{thebibliography}{99}

\bibitem[1]{abd10} A.~A.~Abdo, M.~Ackermann, M.~Ajello, \etal\ \sci, {\bf 329} (2010) 817. %(V407 Cyg)

\bibitem[2]{ack14} M.~Ackermann, M.~Ajello, A.~Albert, \etal\ \sci, {\bf 345} (2014) 554. %(LAT novae)

\bibitem[3]{atw09} W.~B.~Atwood, A.~A.~Abdo, M.~Ackermann, \etal\ \apj\ {\bf 697} (2009) 1071.

\bibitem[4]{ban14} D.~P.~K.~Banerjee, V.~Joshi, V.~Venkataraman, \etal\ \apjl\ 785 (2014) L11. 

\bibitem[5]{bed13} W.~Bednarek, \aph\ {\bf 43} (2013) 81.

\bibitem[6]{che10}  C.~C.~Cheung, D.~Donato, E.~Wallace, \etal\ \atel, {\bf 2487} (2010) 1. 

\bibitem[7]{che12}  C.~C.~Cheung, \Fermi-LAT collaboration, in {\it 4th Fermi Symposium}, eConf {\bf C121028} (2012) 106.    %; arXiv:1304.3475 

\bibitem[8]{che12mon1} C.~C.~Cheung, E.~Hays, T.~Venters, D.~Donato, R.~H.~D.~Corbet, \atel, {\bf 4224} (2012) 1.   %Fermi LAT Detection of a New Gamma-ray Transient in the Galactic Plane: Fermi J0639+0548

\bibitem[9]{che12mon2} C.~C.~Cheung, S.~N.~Shore, I.~De Gennaro Aquino, \etal\ \atel, {\bf 4310} (2012) 1. 
           
\bibitem[10]{che13a} C.~C.~Cheung, P.~Jean, S.~N.~Shore, \atel, {\bf 5649} (2013) 1.
         
\bibitem[11]{che13b} C.~C.~Cheung, P.~Jean, S.~N.~Shore, \atel, {\bf 5653} (2013) 1.
           
\bibitem[12]{che14} C.~C.~Cheung, P.~Jean, S.~N.~Shore, \atel, {\bf 5879} (2014) 1.
           
\bibitem[13]{che15a}  C.~C.~Cheung, P.~Jean, S.~N.~Shore, \atel, {\bf 7283} (2015) 1.
           
\bibitem[14]{che15b} C.~C.~Cheung, P.~Jean, S.~N.~Shore, \atel, {\bf 7315} (2015) 1.
           
\bibitem[15]{cho14} L.~Chomiuk, J.~D.~Linford, J.~Yang, \etal\ \nat, {\bf 514} (2014) 339. 
           
\bibitem[16]{del02} M.~Della Valle, L.~Pasquini, D.~Daou, \& R.~E.~Williams, \aap\ {\bf 390} (2002) 155.
           
\bibitem[17]{fin15} T.~Finzell, L.~Chomiuk, U.~Munari, \& F.~M.~Walter, 2015, submitted; arXiv:1506.04953 
           
\bibitem[18]{fuj12} S.~Fujikawa, IAU CBAT, No.~3202 (2012).   %Nova Monocerotis 2012 = Pnv J06393874+0553520  %International Astronomical Union Central Bureau for Astronomical Telegrams
           
\bibitem[19]{her08} M.~Hernanz,  in {\it Classical Novae}, M.~F.~Bode, A.~Evans, Eds.~Cambridge UP, 2nd Ed.~(2008) 252.
           
\bibitem[20]{hil12} A.~B.~Hill, \Fermi-LAT collaboration, in {\it 4th Fermi Symposium}, eConf {\bf C121028} (2012) 112.
           
\bibitem[21]{ita13} K.~Itagaki, IAU CBAT, No.~3628 (2013).   %Nova Delphinus 2013 = PNV J20233073+2046041
           
\bibitem[22]{joh93} W.~N.~Johnson, R.~L.~Kinzer, J.~D.~Kurfess, \etal\ \apjs\ {\bf 86} (1993) 693.
           
\bibitem[23]{lei99} M.~D.~Leising, J.~E.~Grove, R.~A.~Kroeger, \& J.~D.~Kurfess, in {\it 5th Compton Symposium} (1999), A136.
           
\bibitem[24]{lin15} J.~D.~Linford, V.~A.~R.~M.~Ribeiro, L.~Chomiuk, \etal\ \apj\ {\bf 805} (2015) 136. 
                        
\bibitem[25]{muk01} K.~Mukai, \& M.~Ishida, \apj\ {\bf 551} (2001) 1024.
           
\bibitem[26]{mun13} U.~Munari, S.~Dallaporta, F.~Castellani, \etal\ \mnras\ {\bf 435} (2013) 771. 
           
\bibitem[27]{mun15} U.~Munari, F.~M.~Walter, A.~Henden, \etal\ {\it Information Bulletin on Variable Stars}, {\bf 6139} (2015) 1. 
           
\bibitem[28]{nis10} K.~Nishiyama, F.~Kabashima, IAU CBAT, No.~2199 (2010).
             
\bibitem[29]{ori01} M.~Orio, A.~Parmar, R.~Benjamin, \etal\ \mnras\ {\bf 326} (2001) L13.
           
\bibitem[30]{raz10} S.~Razzaque, P.~Jean, \& O.~Mena, \prd, {\bf 82} (2010) 123012. 
           
\bibitem[31]{sch10} B.~E.~Schaefer, \apjs\ {\bf 187} (2010) 275. 
             
\bibitem[32]{sch14} G.~H.~Schaefer, T.~T.~Brummelaar, D.~R.~Gies, \etal\ \nat, {\bf 515} (2014) 234. 
           
\bibitem[33]{sea13} J.~Seach, IAU CBAT, No.~3732 (2013).
           
\bibitem[34]{sea15} J.~Seach, IAU CBAT, No.~4080 (2015).
           
\bibitem[35]{sho03} S.~N.~Shore, G.~Schwarz, H.~E.~Bond, \etal\ \aj\ {\bf 125} (2003) 1507.
           
\bibitem[36]{sho13} S.~N.~Shore, I.~De Gennaro Aquino, G.~J.~Schwarz, \etal\ \aap\ {\bf 553} (2013) A123. 
           
\bibitem[37]{stu14} R.~Stubbings, IAU CBAT, No.~3803 (2014).
           
\bibitem[38]{tak12} T.~Takahashi, K.~Mitsuda, R.~Kelley, \etal\ \procspie, {\bf 8443} (2012) 84431Z. 
           
\bibitem[39]{tat07} V.~Tatischeff, \& M.~Hernanz, \apjl\ {\bf 663} (2007) L101. 
           
\bibitem[40]{wag12} R.~M.~Wagner, S.~Dont, T.~Bensby, \etal\ \atel, {\bf 4157} (2012) 1.   %MOA 2012 BLG-320: Discovery and Observations of a Nova Candidate Towards the Galactic Bulge
           
\bibitem[41]{wil99} P.~Williams, A.~C.~Gilmore, \iaucirc\ {\bf 7176} (1999) 1.
           
\end{thebibliography}
\end{document}